\documentclass[aps,prl,twocolumn,epsfig,preprintnumbers,superscriptaddress,10pt]{revtex4}

\usepackage{graphicx}
\usepackage{dcolumn}
\usepackage{bm}
\usepackage{epsf,amsfonts,amssymb,epsfig,color}

\newcommand{\be}{\begin{equation}}
\newcommand{\ee}{\end{equation}}
\newcommand{\bea}{\begin{eqnarray}}
\newcommand{\eea}{\end{eqnarray}}
\newcommand{\ba}{\begin{eqnarray}}
\newcommand{\ea}{\end{eqnarray}}

\newcommand{\eqn}[1]{(\ref{#1})}
\newcommand{\beq}{\begin{equation}}
\newcommand{\eeq}{\end{equation}}
\newcommand{\beqa}{\begin{eqnarray}}
\newcommand{\eeqa}{\end{eqnarray}}
\newcommand{\beqar}{\begin{eqnarray*}}
\newcommand{\eeqar}{\end{eqnarray*}}

\def\tr{\rm tr}

\def\pa {\partial}
\def\sac{\, , \,\,\,\,\,}

\def\sqg{\sqrt{g}}

\def\t7 {T_\mt{D7}}

\newcommand{\mq}{M_\mt{q}}      

\newcommand{\mt}[1]{\textrm{\tiny #1}}

\newcommand{\etal}{{\it et al}}

\newcommand{\id}{\mathbb{I}}
\newcommand{\mub}{\mu_\mt{B}}
\newcommand{\mui}{\mu_\mt{I}}

\def\sqg{\sqrt{g}}
\newcommand{\rh}{0}
\newcommand{\mmes}{M_\mt{mes}}

\begin{document}

\preprint{ICCUB-12-075}

\title{Off-diagonal Flavour Susceptibilities from  AdS/CFT}

\author{Jorge Casalderrey-Solana} 
\affiliation{Departament d'Estructura i Constituents
de la Mat\`eria, 
Universitat de Barcelona, Mart\'\i \ i Franqu\`es 1, 08028 Barcelona, Spain
}
\author{David Mateos}
\affiliation{Instituci\'o Catalana de Recerca i Estudis Avan\c cats (ICREA), Barcelona, Spain}

\affiliation{Departament de F\'\i sica Fonamental \&  Institut de Ci\`encies del Cosmos, Universitat de Barcelona, Mart\'{\i}  i Franqu\`es 1, E-08028 Barcelona, Spain}


\begin{abstract}
We study flavour susceptibilities in the ${\cal N}=4$ $SU(N)$ super Yang-Mills plasma coupled to two quark flavours at strong coupling and large $N$ by means of its gravity dual. 
The off-diagonal susceptibility is $1/N$-suppressed and we compute it as a one-loop effect on the gravity side. Contrary to naive extrapolation from perturbative results, it attains a finite value in the infinite-coupling limit. Moreover, its parametric form is independent of whether or not  meson bound states exist in the plasma. We conclude that caution must be exercised when drawing conclusions about the QCD plasma from lattice calculations of quark susceptibilities. 
\end{abstract}

\maketitle

\noindent
{{\bf 1. Introduction.}} 
At a temperature $T_c\approx170$ MeV, Quantum Chromodynamics (QCD) undergoes a cross over into a deconfined phase \cite{Cheng:2009zi}, 
the so-called quark-gluon plasma (QGP). Asymptotic freedom ensures that a description in terms of weakly coupled quarks and gluons is applicable at $T \gg T_c$. An outstanding question is whether a similar description in terms of weakly coupled quasiparticles is possible in the vicinity of $T_c$. Although this regime can be rigorously studied in lattice-regularized QCD, in this formulation the nature of the underlying degrees of freedom is not directly apparent but instead must be inferred from the behaviour of computable quantities. It is thus important to check the validity of these inferences in gauge theory plasmas in which the underlying degrees of freedom are understood both at weak and at strong coupling.

Computable quantities in finite-temperature lattice QCD consist mainly of thermodynamic potentials and charge susceptibilities. Unfortunately, the former are rather insensitive to the nature of the effective degrees of freedom. This is clearly illustrated by the four-dimensional $\mathcal{N}=4$ super Yang-Mills (SYM) plasma, which has proven to be a good toy model for the QGP just above $T_c$ (see e.g.~\cite{CasalderreySolana:2011us} and references therein). At weak coupling the SYM plasma is well described as a gas of gluons, adjoint scalars and fermions. At strong coupling no quasiparticles exist and the effective description involves black hole in five-dimensional anti-de Sitter (AdS) space \cite{witten}. Despite this dramatic difference, in the infinite-coupling limit the pressure saturates at 75\% of the free value 
\cite{Gubser:1996de}. Remarkably, the pressure of QCD in the region of interest deviates from the free limit by a comparable fraction \cite{Cheng:2009zi}. 
In addition, perturbative calculations generically suffer from problems when extrapolated to such low temperatures and the theoretical uncertainties become comparable to the deviations above  \cite{Blaizot:2000fc}. 

In contrast to thermodynamic potentials, the flavor structure of the QGP has been regarded as a better discriminant for the presence of quasiparticles. Quark susceptibilities are defined as 
\be
\label{chigendef}
\chi_{ij} = \left. \frac{T}{V} \frac{\pa^2 \log Z}{\pa \mu_i \pa \mu_j} \right |_{\mu_i=0} \, ,
\ee
with $Z$ the QCD partition function and 
$\mu_i=\mu_{u,d}$ the $u$ and $d$ chemical potentials.
Early lattice calculations of diagonal quark susceptibilities \cite{Cheng:2008zh} indicated a rapid convergence 
to the free-theory limit above $T_c$. This were contrasted with a relatively larger deviation (by a factor of 2) \cite{contrast} between the free and the strongly coupled limits of the $\mathcal{N}=4$ SYM plasma.  
However, more recent lattice results 
\cite{Borsanyi:2011bm,Bazavov:2011nk} 
show a slower approach to the free limit, with sizable deviations up to temperatures of $2 \, T_c$.  As before, in this regime the uncertainty arising from perturbative  calculations 
\cite{Blaizot:2001vr} 
 becomes comparable to the difference between the free and the strongly coupled limits \cite{Borsanyi:2011bm}.

Similarly, off-diagonal quark susceptibilities have also been proposed \cite{Koch:2005vg} as a sensitive probe of the degrees of freedom of the plasma, based on the assumption that a dominance of quark-like quasiparticles must lead to uncorrelated flavors and hence to small off-diagonal susceptibilities. In perturbative $SU(N)$ QCD, the off-diagonal susceptibility is strongly suppressed with respect to the diagonal one because \cite{Blaizot:2001vr}
\be
\label{chipert}
\frac{\chi_{u d}}{\chi_{u u}} = -\frac{(N^2-1)(N^2-4)}{128 \pi^6 N^2} \frac{\lambda^3}{N^3} \log \frac{1}{g} \, ,
\ee
with $g$ the coupling constant and $\lambda=g^2 N$ the `t Hooft coupling. In contrast, lattice results have shown that 
this ratio is sizable up to temperatures as high as $2\, T_c$. Since this cannot be fully accounted for by the perturbative result \eqn{chipert}, it has been taken as an indication of the presence of bound  states in the plasma \cite{Ratti:2011au} (see also \cite{Shuryak:2004tx}). In contrast to the previous cases, this inference cannot be tested in strongly coupled $\mathcal{N}=4$ SYM coupled to dynamical quarks because the off-diagonal quark susceptibility in this theory has not been calculated. The reason is that, as can be seen from  \eqn{chipert}, the off-diagonal susceptibility is $1/N$-suppressed in the large-$N$ limit, and hence it corresponds to a quantum effect on the dual gravity side. The purpose of this Letter is to perform this calculation and thus  shed light on the interpretations of the lattice results described above. We will focus on determining the qualitative form of the result as well as its leading parametric dependence.

\noindent
{{\bf 2. Gravity calculation.}}
The $\mathcal{N}=4$ SYM  plasma coupled to two quark flavours ($u$ and $d$) of equal masses is dual to two coincident D7-brane probes \cite{karchkatz} in $AdS_5 \times S^5$ in the presence of a black hole horizon \cite{witten}. At temperatures $T< M$ the branes sit entirely outside the horizon in the so-called `Minkowski phase' (MP) \cite{Babington:2003vm}. The spectrum in the flavour sector consists of mesons with masses $\mmes\sim M$ and parametrically heavier quarks with masses 
$\mq \sim \sqrt{\lambda} \, M$ \cite{Kruczenski:2003be}.
On the gravity side the quarks correspond to long strings stretching between the branes and the horizon, and the mesons to normal-mode fluctuations of the branes. Both sets of excitations are good quasiparticles since their widths vanish in the limit $N, \lambda \to \infty$. In contrast, at temperatures $T > M$ the branes pierce the horizon in the so-called `black hole phase' (BHP). In this phase no quasiparticles exist. Meson-like excitations are now broad resonances whose frequencies posses large imaginary parts, and correspond on the gravity side to quasinormal modes on the brane \cite{qnm}. Quarks are effectively massless in this phase, but the widths of quark-like excitations are comparable to their energies. We will determine the susceptibilities in both phases. 

The embedding of the D7-branes in $AdS_5 \times S^5$ in the presence of the black hole depends only on the dimensionless ratio $M/T$. Following \cite{CasalderreySolana:2010xh}, we write the induced metric on the branes (in Euclidean signature) as $ds^2=L^2 ds^2(g)$, with 
 \be
ds^2 (g) = g_{\tau \tau} d \tau^2 + g_{xx} dx_i^2 
+ g_{rr} dr^2 + g_{\Omega\Omega} d \Omega^2_3 \,.
\label{g}
\ee
The  gauge theory coordinates $x=\{\tau, x^i\}$ and the AdS radial coordinate $r$ are dimensionless. Dimensions can be restored through the replacements $x \to \pi T x$, $r \to r/\pi L^2 T$. The last term in \eqn{g} is the metric on the $S^3$ wrapped by the branes. The metric functions depend on $M/T$ and $r$, which we choose so that $r\in (0,\infty)$ in both phases. In the BHP $g_{\tau \tau}$ vanishes at the horizon, i.e.~$g_{\tau \tau}=0$ at $r=0$.

The full action describing the non-Abelian $U(2)$ gauge theory on the branes is not known. However, for our purposes it suffices to consider the Yang-Mills action \cite{else}
\be
\label{ymac}
S_\mt{E}=\frac{1}{4 e^2} \int dr d^4 x \sqrt{g}\,\mbox{Tr} \left\{ 
F^{\mu\nu} F_{\mu\nu}\right\} \,,
\ee
where the trace is over flavour indices and $\sqrt{g}$ includes $g_{\Omega\Omega}$. The effective coupling constant is
$e^2= 4\pi^4/N$ \cite{CasalderreySolana:2010xh}, so the large-$N$ limit corresponds to the classical limit.

We introduce chemical potentials by turning on a background value for the gauge field on the branes  
\be
\label{Bgen}
B_\tau= -i \left(\mub \id+ \mui \sigma^3 \right) H(r) \, ,
\ee
where $\mub=(\mu_u+\mu_d)/2 \pi T$ and $\mui=(\mu_u-\mu_d)/2 \pi T$ are the (dimensionless) baryon and isospin chemical potentials, repectively, and $\sigma^3$ is the two-dimensional Pauli matrix. The  function $H(r)$ is obtained by solving the classical equations, which yields $H(r)=1$ in the MP and 
\be
H(r)=1- f {\textstyle \left( \frac{M}{T} \right)}
\int^\infty_r dr \, \frac{g_{rr} g_{\tau\tau}}{\sqrt{g}} 
\ee
in the BHP, where 
\be
f^{-1}{\textstyle \left( \frac{M}{T} \right)}= \int_{\rh}^\infty dr \, \frac{g_{rr} g_{\tau\tau}}{\sqrt{g}} 
\ee
is a dimensionless function that depends only on the embedding of the branes. Substituting  (\ref{Bgen}) into the action (\ref{ymac}) we obtain (the potential-dependent part of) the free energy, defined as $F=TS_\mt{E}=- T\log Z$. The result is $F_0=0$ in the MP and  
 \be
 \label{f0}
 F_0=-V \,f {\textstyle \left( \frac{M}{T} \right)} \, \frac{\pi^2 T^2 }{2 e^2} \left(\mu^2_u+ \mu^2_d\right)  
 \ee
in the BHP, where the superscript `0' indicates that this is the classical result. Using \eqn{chigendef} we get the susceptibilities: 
\bea
&& \chi_{ud}^0=0 \sac \chi_{uu}^0 = \chi_{dd}^0 = 0 \,,
\qquad \,\,\,  \,\,\,\,\,\,\,\,\,\,\,\,\,\,\,\,\,\,\,\,\,\,\,\,\,\,
\mbox{[MP]} \,\,\,\,\,\,\,\,\,\, \\
&& \chi_{ud}^0=0 \sac 
\chi_{uu}^0 = \chi_{dd}^0 = {\textstyle \frac{1}{4\pi^2}}
f {\textstyle \left( \frac{M}{T} \right)} T^2 N \,.
\,\,\,\,\, \,\,\,\mbox{[BHP]} \,\,\,\,\,\,\,\,\,\,
\label{quarks}
\eea
The diagonal values reproduce the original calculation \cite{Kobayashi:2006sb}. We will come back to these results in Sec.~3. Suffice it to note here that the off-diagonal susceptibilities vanish at leading order in $N$ in both phases, as expected. We will compute the 
leading-order correction in the $1/N$ expansion. The corrections  that are sensitive to the chemical potentials arise from  quantum fluctuations of the brane fields around the classical solution  (\ref{Bgen}). For concreteness  we will focus on vector fluctuations, i.e.~we set
\be
\label{defa}
A_\mu= B_\mu + \frac{e}{\sqrt{2}}  \, a_\mu \,.
\ee
The matrix-valued  field $a$ can be decomposed into modes with different charges under the $U(2)$ flavour group as 
\be
a_\mu= a^{\mt{B}}_\mu \, \id + a^a_\mu \,\sigma^a  \,.
\ee
The Lagrangian $\mathcal{L}$ for these excitations is obtained by substituting  eqn.~(\ref{defa}) into the action (\ref{ymac}).  To capture the leading $1/N$ corection we expand the lagrangian to leading order in $e$, i.e.~to quadratic order in $a$. Due to the commutation relations among Pauli matrices, the only components that couple to the background field \eqn{Bgen} are $a_\mu^{1,2}$. For simplicity, henceforth we restrict ourselves to the two transverse excitations that satisfy $k^i a_i^{1,2}=0$ in Fourier space. Since the Euclidean time  is periodic under $\tau \rightarrow \tau + \pi$, each of these can be expanded as $a^a=\sum_\ell e^{-2\ell i} a^a_\ell$, so we finally arrive at 
\bea
\mathcal{L}&=&\sum_\ell\sqrt{g} g^{xx} \left[ ( 4 \ell^2g^{\tau \tau}  + k^2 g^{xx}) a^*_\ell a_\ell +g^{rr} \partial_r a^*_\ell \pa_r a_\ell\right] 
\nonumber
\\ 
&&
\hspace{0.5cm}
-\sqrt{g} g^{\tau \tau} g^{xx} \left[ 8 i \ell \mui H + 
4 \mui^2 H^2  \right] a^*_\ell a_\ell \,,
\eea
where we have introduced $a_\ell=\left(a^1_\ell+i\,a^2_\ell \right)/\sqrt{2}$.  For Minkowski embeddings, for which $H(r)=1$, this is  the Lagrangian of a scalar with a chemical potential $\mui $. 
 
To determine the partition function, we will express a general fluctuation $a$ in term of a basis of (radial) functions obtained by solving the eigen-value problem 
 \be
 \label{basis}
 -\frac{1}{\sqg  (g^{xx})^2}\pa_r\Big[
 \sqg g^{rr} g^{xx} \pa_r \phi_{\ell n}  \Big] + 
4\ell^2 \frac{g^{\tau \tau}}{ g^{xx}}  \phi_{\ell n}= 
\kappa^2_{\ell n}  \phi_{\ell n} \,.
\ee
Normalizability requires the eigen-functions $\phi_{\ell n}$ to vanish as $r\to \infty$. As $r\to0$ regularity requires $\pa_r \phi_{\ell n}  \to 0$ in the MP \cite{Kruczenski:2003be}, and $\phi_{\ell n}  \to r^{2|\ell |}$ in the BHP \cite{Denef:2009kn}. In addition, these functions obey the orthonormality conditions
\be
\label{normR}
\int^\infty_0 d r \sqg  (g^{xx})^2 \phi_{\ell n} 
\phi_{\ell m }= \delta_{nm} \,.
\ee
 In this basis, the $\mui$-independent part of the Lagrangian is diagonal, which yields a Gaussian partition function for 
 $\mui=0$ given by 
 \be
 \label{zfree}
Z(0) = \prod_{\ell n k} \int d c^*_{\ell n}  dc_{\ell n} \exp 
\left[ - c^*_{\ell n} c_{\ell n} \left( k^2+\kappa^2_{\ell n}\right) \right] \,.
\ee
 The full partition function is obtained by averaging the $\mui$-dependent part over the partition function (\ref{zfree}):
 \bea
 Z(\mui)&=&Z(0) \times \\
&& \hspace{-1.5cm}
 \Big< \exp \Big[ 
V \int \frac{d^3 k}{(2\pi)^3} \sum_{\ell nm} 
c^*_{\ell n} c_{\ell m}
\left(i 8\ell \mui M^{(1)}_{\ell nm} + 4\mui^2 M^{(2)}_{\ell nm}\right)
\Big] \Big> \,, \nonumber
\eea
 where $V$ is the 3-dimensional volume and
 \bea
 \label{s1def}
M^{(1)}_{\ell nm} &=& \int dr  \sqrt{g} \, g^{\tau \tau} g^{x x} H(r)  \phi^*_{\ell n} \phi_{\ell m}\, ,
\\
\label{s2def}
M^{(2)}_{\ell nm} &=& \int dr  \sqrt{g}\,  g^{\tau \tau} g^{x x} H(r)^2  \phi^*_{\ell n} \phi_{\ell m} \,.
\eea
Since we are only interested in susceptibilities, we can consider infinitesimal values of $\mui/T$ and determine the partition function via perturbation theory. Since at vanishing 
$\mui$ the plasma is isospin neutral, the leading-order perturbation in $\mui/T$ is given by
\bea
\label{dlogZgen0}
\delta \log Z&=& \mui^2
 \, V \int \frac{d^3 k}{(2\pi)^3} \sum_l \sum_n \\ 
&&
 \left[  \frac{ 4 M^{(2)}_{\ell nn}}{\kappa^2_{ln} + k^2}-\sum_{m}\frac{ 32 \ell^2\, M^{(1)}_{\ell nm} M^{(1)}_{\ell mn}}{(\kappa^2_{\ell n} + k^2)(\kappa^2_{\ell m} + k^2)}\right] \,.\nonumber
\eea
This expression only depends on $\mui$ and on the mode spectrum, which is determined by the branes' embedding, i.e.~by $M$ and $T$ alone. In particular, it is both $\lambda$- and $N$-independent, irrespectively of the phase. Taking derivatives with respect to $\mu_d$ and $\mu_u$ we conclude that the contribution of the transverse vector modes takes the form
\be 
\chi_{uu}^1=\chi_{dd}^1=-\chi_{ud}^1=
h  {\textstyle \left( \frac{M}{T} \right)} T^{2} \,, \,\,\,\,\,\,\,
\mbox{[MP\&\,BHP]}
\label{both} 
\ee
where $h$ is a coupling- and $N$-independent function, and the superscript `1' indicates that this is the one-loop result. We will come back to it in Sec.~3.

The interpretation of eqn.~(\ref{dlogZgen0}) is obscured by the fact that the  Euclidean solutions (\ref{basis}) do not correspond to the physical excitations of the plasma, i.e.~to  mesons in the MP or to wide resonances (quasinormal modes) in the BHP. Both of these are defined in Lorentzian signature, and they are characterized by dispersion relations $\omega_n (k)$ and the associated radial wave-functions $\varphi_n(k)$.
The connection can be made clear by noticing that  the Euclidean differential equation (\ref{basis}) and the corresponding boundary conditions are obtained from their Lorentzian counterparts upon the analytic continuation 
\be 
k\rightarrow \pm i \kappa_{\ell n} \sac 
\omega \rightarrow -2 i \ell \,. 
\ee
This implies that, viewed as functions of complex momentum, the dispersion relations obey $\omega_n ( \pm i\kappa_{\ell n}) = -2i\ell$, and similarly that 
$\varphi_n( \pm i \kappa_{\ell n})= \phi_{\ell n}$. Followig 
 \cite{Denef:2009kn} we rewrite eqn.~(\ref{dlogZgen0}) by noting that its poles coincide with those of 
\bea
\label{dlogZgen}
\delta \log Z&=& \mui^2 \, V 
\int \frac{d^3 k}{(2\pi)^3} \sum_\ell \sum_n \\ 
&&
\hspace{-1.5cm} \left[ \frac{ 4 e^{P^{(2)}_n(k)}
N^{(2)}_{nn}(k) }{\left| i \omega_{n}(k)-2\ell \right|^2}-\sum_{m}\frac{ 32 \,\ell^2\, e^{P^{(1)}_{nm}(k)} 
N^{(1)}_{nm} (k) N^{(1)}_{mn}(k) }{\left|i \omega_{n}(k)-2\ell \right|^2 \left| i \omega_{m}(k)-2\ell \right|^2}\right] \,, \nonumber
\eea
where $P^{(2)}_n(k)$ and $P^{(1)}_{n m}(k)$ are 
polynomial functions and
 \bea
 \label{n1}
N^{(1)}_{nm} (k) &=& \int dr  \sqrt{g} \, g^{\tau \tau} g^{x x} H(r)  \varphi^*_{n}(k,r) \varphi_{n}(k,r)\, ,
\\ 
\label{n2}
N^{(2)}_{nm} (k) &=& \int dr  \sqrt{g} \, g^{\tau \tau} g^{x x} H(r)^2  \varphi^*_{n}(k,r) \varphi_{n}(k,r) \,\,\,\,\,\,\,
\eea
have no poles in the complex plane. 

The coefficients \eqn{n1}-\eqn{n2} now have a simple interpretation. The radial wave functions of the physical excitations satisfy the orthonormality conditions (different from (\ref{normR})) 
 \be
 \int^\infty_0 d r \sqg  \, g^{\tau \tau} g^{xx} 
\varphi^*_{n}(k,r) \varphi_{m}(k,r) = \delta_{nm}  \,.
 \ee
Therefore the functions $N^{(q)}_{nm}(k)$ are the matrix elements of the operator $H(r)^q$ in this basis. It is clear that these mixing coefficients determine the (second order) perturbation of the physical modes in the presence of the external field. 

In the MP the function $H(r)$ is unity and hence $N^{(q)}_{nm}(k)=\delta_{nm}$. Also, in this case it is easy to see \cite{else} that the polynomials $P^{(i)}(k)$ vanish. Since the frequency of the modes does not depend on the Matsubara frequencies $2\ell$, eqn.~(\ref{dlogZgen}) coincides with that of a charged bosonic field. Performing the sum over $\ell$ we then arrive at the one-loop contribution from transverse vector mesons
\be
\chi_{ud}^1=\sum_n -2\, \frac{2}{T} \int \frac{d^3 k}{(2\pi)^3} \frac{e^{-\omega_n(k)/T}}{\left(1-e^{-\omega_n(k)/T}\right)^2}\,, 
\,\,\, \mbox{[MP]} 
\label{nomix}
\ee
where we have restored physical units and one factor of 2 arises from the two transverse polarizations. We conclude that in the MP there is no mode mixing, i.e.~mesons contribute to the partition function independently of one another. In contrast, in the BHP 
$H(r)$ is not constant and the mixing coefficients $N^{(q)}_{nm}$ are not diagonal. As a consequence, the contributions  to the one-loop susceptibilities of the broad resonances present in the plasma mix with one another. 

\noindent
{{\bf 3. Discussion.}} The ${\cal N}=4$ SYM plasma provides a rich model in which several expectations about the behaviour of  quark susceptibilities at strong coupling can be tested. One important conclusion is that, contrary to what naive extrapolation of the perturbative formula \eqn{chipert} might suggest, the off-diagonal susceptibility attains a finite value in the limit $\lambda \to \infty$, as illustrated by eqn.~\eqn{both}. On the gravity side this follows simply from the overall factor $1/e^2$ in the action (\ref{ymac}), together with the quadratic nature of the leading-order Lagrangian for the fluctuations. Since these features are common to all brane fluctuations (bosonic or fermionic), 
$\chi_{ud}$ will remain $\lambda$-independent once the contributions  of all the fluctuations are included. This conclusion applies regardless of the nature of the effective degrees of freedom in the plasma. Indeed, the MP contains quark-like and meson-like quasiparticles, whereas no quasiparticle excitations of either type exist in the BHP. Despite this dramatic difference, the off-diagonal susceptibilities are parametrically the same in the two phases. In particular, the BHP illustrates the fact that non-vanishing off-diagonal susceptibilities do not imply the existence of narrow meson-like states. This limited sensitivity suggests that caution  should be exercised when extracting conclusions about the value of the coupling constant or about the nature of the underlying degrees of freedom in the QGP from lattice calculations of the off-diagonal susceptibilities. 
 
The fact that the off-diagonal susceptibility is of order $N^0$ in both phases is due to the fact it arises from contributions of meson-like states in both cases. In the MP these are narrow quasiparticles and eqn.~\eqn{nomix} shows that each of them contributes independently, as expected \cite{Koch:2005vg}. In the BHP these are broad resonances and their contributions mix with one another. 

Diagonal susceptibilities are of order $N^0$ in the MP and of order $N$ in the BHP. In the MP this is due to the fact that quark-like excitations are parametrically heavier than the mesons by a factor of $\sqrt{\lambda}$, so their contribution to the susceptibilities is not visible at leading order. In contrast, in the BHP quarks are effectively massless and they produce an order-$N$  contribution to the diagonal susceptibilities, eqn.~\eqn{quarks}, despite the fact that they do not give rise to narrow quasiparticles. In the BHP eqn.~\eqn{both} provides the oder-one meson correction to the leading quark contribution.

The change in scaling of the diagonal susceptibilities produces a change in the ratio $\chi_{ud}/\chi_{uu}$ from order one in the MP to order $1/N$ in the BHP. We stress that this strong decrease  in the ratio is not due to a reduction of cross-flavour correlations across the transition but rather to the appearance of additional, flavour-charged degrees of freedom (the quarks) in the BHP.

\noindent
{{\bf Acknowledgments.}}
We are grateful to V.~Koch, J.~Liao and A.~V.~Ramallo for useful discussions. JCS is supported by a RyC 
fellowship. We acknowledge  support from grants 2009-SGR-502, 
2009-SGR-168, MEC FPA2010-20807-C02-01, MEC FPA2010-20807-C02-02, and CPAN CSD2007-00042 Consolider-Ingenio 2010.


\begin{thebibliography}{99}

\bibitem{Cheng:2009zi} 
  M.~Cheng {\it et al.},
  Phys.\ Rev.\ D {\bf 81}, 054504 (2010);
  S.~Borsanyi {\it et al.},
  JHEP {\bf 1011}, 077 (2010).
  


\bibitem{CasalderreySolana:2011us}
  J.~Casalderrey-Solana {\it et al}, 
  [arXiv:1101.0618 [hep-th]].

  \bibitem{witten} 
  E.~Witten,
  Adv.\ Theor.\ Math.\ Phys.\  {\bf 2}, 505 (1998).

\bibitem{Gubser:1996de} 
  S.~S.~Gubser, I.~R.~Klebanov and A.~W.~Peet,
  Phys.\ Rev.\ D {\bf 54}, 3915 (1996).
  
\bibitem{Blaizot:2000fc} 
  J.~P.~Blaizot, E.~Iancu and A.~Rebhan,
  Phys.\ Rev.\ D {\bf 63}, 065003 (2001);
  %
  A.~Hietanen {\it et al}, 
  Phys.\ Rev.\ D {\bf 79}, 045018 (2009).
  


\bibitem{Cheng:2008zh}
  M.~Cheng, P.~Hendge, C.~Jung, F.~Karsch, O.~Kaczmarek, E.~Laermann, R.~D.~Mawhinney, C.~Miao {\it et al.},
  Phys.\ Rev.\  {\bf D79}, 074505 (2009).
  [arXiv:0811.1006 [hep-lat]].
  
   \bibitem{contrast} 
  S.~S.~Gubser,
  Nucl.\ Phys.\ B {\bf 551}, 667 (1999);
  D.~T.~Son and A.~O.~Starinets,
  JHEP {\bf 0603}, 052 (2006).

  
\bibitem{Borsanyi:2011bm}
  S.~Borsanyi {\it et al} [ Wuppertal-Budapest Collaboration ],
  [arXiv:1109.5030 [hep-lat]].
  %
\bibitem{Bazavov:2011nk}
  A.~Bazavov {\it et al}, 
  [arXiv:1111.1710 [hep-lat]];
  S.~Mukherjee,
  [arXiv:1107.0765 [nucl-th]];
  %
  S.~Borsanyi \etal
  arXiv:1112.4416 [hep-lat].
   
     
     
\bibitem{Blaizot:2001vr}
  J.~P.~Blaizot, E.~Iancu and A.~Rebhan,
  Phys.\ Lett.\  B {\bf 523}, 143 (2001);
  %
  A.~Vuorinen,
  Phys.\ Rev.\ D {\bf 67}, 074032 (2003).
     
\bibitem{Koch:2005vg}
  V.~Koch, A.~Majumder, J.~Randrup,
  Phys.\ Rev.\ Lett.\  {\bf 95}, 182301 (2005).
  
  \bibitem{Ratti:2011au}
  C.~Ratti \etal, 
  [arXiv:1109.6243 [hep-ph]].
  
\bibitem{Shuryak:2004tx}
  E.~V.~Shuryak, I.~Zahed,
  Phys.\ Rev.\  {\bf D70}, 054507 (2004).
  
    \bibitem{karchkatz} 
  A.~Karch and E.~Katz,
  JHEP {\bf 0206}, 043 (2002).


  
  \bibitem{Babington:2003vm} 
  J.~Babington \etal, 
  Phys.\ Rev.\ D {\bf 69}, 066007 (2004);
%
  M.~Kruczenski \etal, 
  JHEP {\bf 0405}, 041 (2004);
%
  I.~Kirsch,
  Fortsch.\ Phys.\  {\bf 52}, 727 (2004);
  %
  D.~Mateos, R.~C.~Myers and R.~M.~Thomson,
  Phys.\ Rev.\ Lett.\  {\bf 97}, 091601 (2006); 
  JHEP {\bf 0705}, 067 (2007).
 

\bibitem{Kruczenski:2003be} 
  M.~Kruczenski \etal, 
  JHEP {\bf 0307}, 049 (2003).

\bibitem{qnm} 
  C.~Hoyos-Badajoz, K.~Landsteiner and S.~Montero,
  JHEP {\bf 0704}, 031 (2007).

\bibitem{CasalderreySolana:2010xh} 
  J.~Casalderrey-Solana, D.~Fernandez and D.~Mateos,
  Phys.\ Rev.\ Lett.\  {\bf 104}, 172301 (2010);
  JHEP {\bf 1011}, 091 (2010).

\bibitem{else} 
J.~Casalderrey-Solana and D.~Mateos, ``Flavour susceptibilities from quantum effects in anti-de Sitter space", in preparation.

\bibitem{Kobayashi:2006sb} 
  S.~Kobayashi \etal, 
  JHEP {\bf 0702}, 016 (2007).



\bibitem{Denef:2009kn} 
  F.~Denef, S.~A.~Hartnoll and S.~Sachdev,
  Class.\ Quant.\ Grav.\  {\bf 27}, 125001 (2010).





\end{thebibliography}
\end{document}